\documentclass[aps,prl,twocolumn,showpacs,superscriptaddress,groupedaddress]{revtex4}  % for review and submission
\usepackage{graphicx}  % needed for figures
\usepackage{dcolumn}   % needed for some tables
\usepackage{bm}        % for math
\usepackage{amssymb}   % for math
\usepackage{color}
\usepackage[colorlinks,]{hyperref}
\usepackage{mathrsfs}
\usepackage{amsfonts}
\usepackage{amsthm}
\usepackage{dcolumn}
\usepackage{txfonts}
\begin{document}

% The following information is for internal review, please remove them for submission
\widetext
%\leftline{Version xx as of \today}
%\leftline{Primary authors: Joe E. Physics}
%\leftline{To be submitted to (PRL, PRD-RC, PRD, PLB; choose one.)}
%\leftline{Comment to {\tt d0-run2eb-nnn@fnal.gov} by xxx, yyy}
%\centerline{\em D\O\ INTERNAL DOCUMENT -- NOT FOR PUBLIC DISTRIBUTION}

% the following line is for submission, including submission to the arXiv!!
%\hspace{5.2in} \mbox{Fermilab-Pub-04/xxx-E}

\title{Band-selective shaped pulse for high fidelity quantum control in diamond} %Title of paper

\author {Yan-Chun Chang}
\author{Jian Xing}
\affiliation{Beijing National Laboratory for Condensed Matter Physics, Institute of Physics, Chinese Academy of Sciences, Beijing 100190, China}
\author {Fei-Hao Zhang}
\affiliation{Tsinghua National Laboratory for Information Science and Technology, Beijing 100084, China}
\affiliation{State Key Laboratory of Low-Dimensional Physics and Department of Physics, Tsinghua University, Beijing 100084, P R China}
\author{Gang-Qin Liu}
\author{Qian-Qing Jiang}
\affiliation{Beijing National Laboratory for Condensed Matter Physics, Institute of Physics, Chinese Academy of Sciences, Beijing 100190, China}
\author {Wu-Xia Li}
\affiliation{Beijing National Laboratory for Condensed Matter Physics, Institute of Physics, Chinese Academy of Sciences, Beijing 100190, China}
\author{Chang-Zhi Gu}
\affiliation{Beijing National Laboratory for Condensed Matter Physics, Institute of Physics, Chinese Academy of Sciences, Beijing 100190, China}
\affiliation{Collaborative Innovation Center of Quantum Matter, Beijing 100871, China.}
\author {Gui-Lu Long}
\affiliation{Tsinghua National Laboratory for Information Science and Technology, Beijing 100084, China}
\affiliation{State Key Laboratory of Low-Dimensional Physics and Department of Physics, Tsinghua University, Beijing 100084, P R China}
\affiliation{Collaborative Innovation Center of Quantum Matter, Beijing 100871, China.}
\author{Xin-Yu Pan,}
\email{xypan@aphy.iphy.ac.cn}
\affiliation{Beijing National Laboratory for Condensed Matter Physics, Institute of Physics, Chinese Academy of Sciences, Beijing 100190, China}
\affiliation{Collaborative Innovation Center of Quantum Matter, Beijing 100871, China.}

\date{\today}

\begin{abstract}
High fidelity quantum control over qubits is of crucial importance
for realistic quantum computing, and it turns to be more
challenging when there are inevitable interactions among qubits. By
employing a band-selective shaped pulse, we demonstrate a high
fidelity flip over electron spin of nitrogen-vacancy (NV) centers in
diamond. In contrast with traditional rectangular pulses, the shaped
pulse has almost equal excitation effect among a sharply edged region
(in frequency domain). So the three sub-levels of host $^{14}$N
nuclear spin can be flipped accurately at the same time, while the
redundant flip of other sublevels (e. g. of a nearby $^{13}$C
nuclear spin ) is well suppressed. The shaped pulse can be applied
to a large amount of quantum systems in which band-selective operation
are required.
\end{abstract}

 \maketitle

%\section{\label{sec:level1}First-level heading}
% sections are not used for PRL papers
\section{Introduction}

High fidelity coherent manipulation is an essential asset for many applications
such as quantum information process (QIP), quantum metrology and quantum sensing.
In physical quantum computation, accurate quantum gates
are required to satisfy the DiVincenzo criteria \cite{DiVincenzo2000}.
Two-bit quantum gate is necessary to completely characterize a quantum process \cite{Poyatos1997}.
More coherently interacting qubits are required for scalable QIP
\cite{cirac2000scalable,helmer2009cavity}.
With local control gate, every qubit need to be precisely addressed and coherent controlled individually
\cite{burgarth2010scalable}.
In principle, each qubit can be distinguished by different spacial locations \cite{yavuz2006fast}
or resonant frequencies
\cite{rippe2005sub}.
However, the control field (for example, microwave) usually cant't be focused to certain qubit without disturbing others \cite{Dolde2014}.
Precise control of individual qubit becomes challenging
because of unwanted interactions \cite{viola1999universal}
or off-resonant excitation \cite{rabi1937space,scully1997quantum}.
Under these circumstances, the fidelity of traditional control decreases obviously.
As a result,
many techniques have been developed to optimal the quantum control fidelity,
such as GRAPE (gradient ascent pulse engineering) algorithm \cite{khaneja2005optimal} and shaped pulse\cite{friedrich1987shaped}.

Shaped pulses is the technique using specially tailored pulses
to suppress the effects of unwanted interaction and off-resonant excitation \cite{mccoy1992selective}.
Both the amplitudes and the phases of the pulse sequences can be modulated
by numerical optimization or analytical approach to achieve the ideal quantum operation.
This methodology has been widely applied in the experiments of NMR (nuclear magnetic resonance) systems \cite{vandersypen2005nmr}.
Similar schemes can also been founded in the laser spectroscopy \cite{bartels2000shaped,tannor1985control}%,rabitz2000whither}.
Furthermore, various pulse shapes, such as the Gaussian shape \cite{bauer1984gaussian},
the Hermite shape \cite{warren1984effects} and the BURP (band-selective, uniform response, pure-phase) family of pulses \cite{geen1991band},
have been elaborated designed for diverse specific applications.
With many outstanding properties, such as the frequency selectivity \cite{geen1991band},
wide-band but sharply edged transition range, self-refocusing behavior \cite{emsley1989self},
robustness to experimental imperfections and universality,
these shaped pulses have depicted excellent performance from various aspects.
It can be used for high fidelity manipulation in various physical systems.

Spin or pseudospin qubits in solid state systems are one promising candidate class for QIP,
such as quantum dots \cite{loss1998quantum} and superconductor circuits\cite{clarke2008superconducting}.
As one of the possible solid state system, Nitrogen-vacancy (NV) center in diamond have drawn much attentions these years
\cite{dolde2013room,Dolde2014,Waldherr2014,
dutt2007quantum,bernien2013heralded}.
Electron spin of NV center can be optically initialized, readout, and be manipulated by microwave pulses at room temperature.
Multi-qubit in NV system are comprised by the hyperfine interaction between NV electron spin and nuclear (ambient carbon nuclei or host nitrogen nucleus) spins \cite{dutt2007quantum} or contiguous electron spins \cite{dolde2013room}.
However, as qubit number increases, the hyperfine interaction spectra become dense \cite{Dolde2014}.
Unwanted interactions and off-resonant excitation make control fidelity decrease.
To overcome the problems,
many pulse optimal techniques have been used
in various physical systems, for example NMR \cite{tovsner2009optimal}, trapped ions \cite{blatt2008entangled} and also solid state systems \cite{dicarlo2009demonstration,Dolde2014}.
In this paper, we demonstrate another technique,
REBURP (refocusing
%band-selective, uniform response, pure-phase. one type of shaped pulse
BURP) pulse,
to optimal control fidelity of the NV center spin.

\section{Rectangular and REBURP pulse}

\begin{figure}
\includegraphics[width=\linewidth]{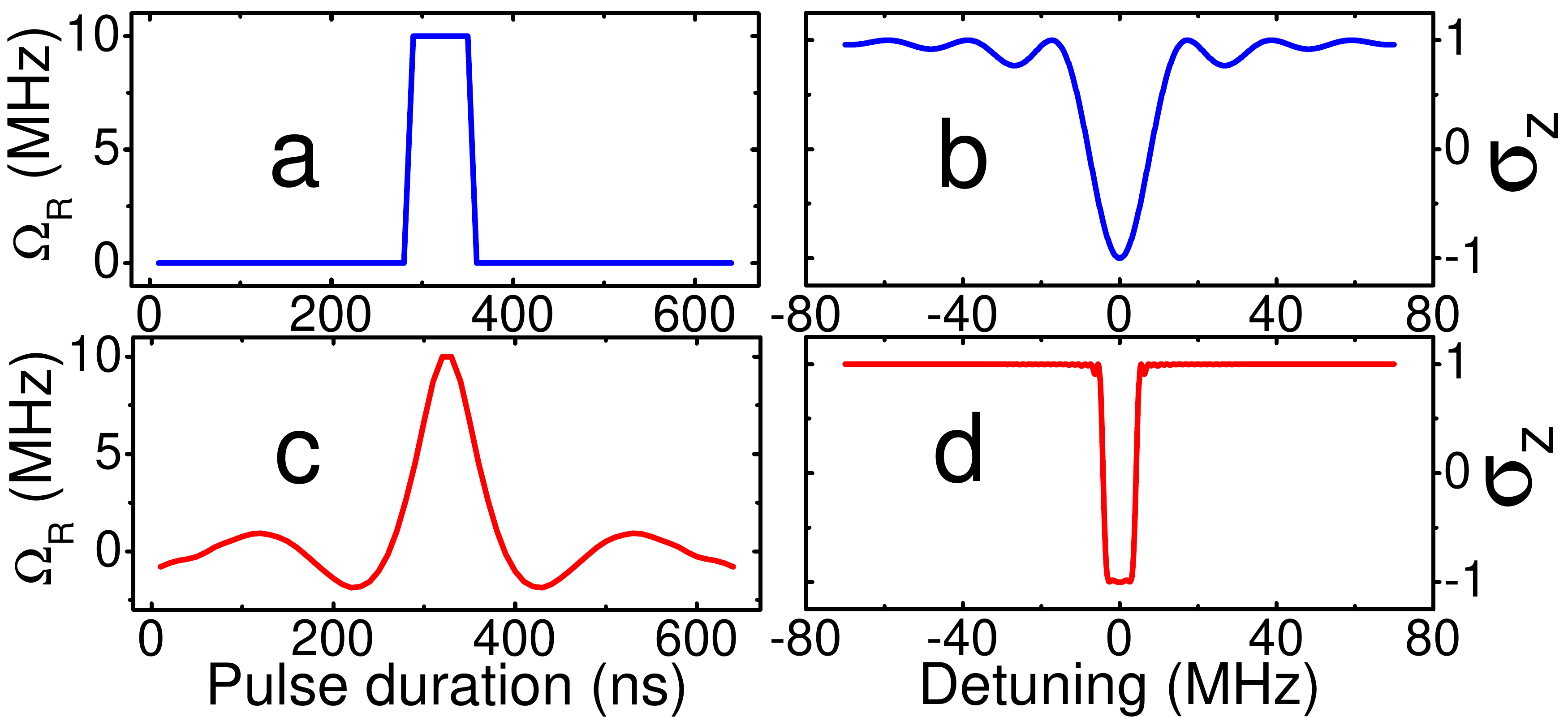} \caption{\label{Shape}
\textbf{Profile and spin response of the rectangular and shaped pulse} (Color online).
(a) A rectangle pulse in time domain.
(b) Spin response (frequency domain) of rectangle $\pi$ pulse.
(c) The REBURP 180 shaped pulse in time domain.
(d) The spin response (frequency domain) of the REBURP 180 pulse.}
\end{figure}

As traditional control signal, rectangular pulses have a wide spectrum in frequency domain,
with many other frequency elements besides
the carrier frequency.
The profile (time domain) and spin response (frequency domain) of rectangle pulse are shown in Fig.\ref{Shape}(a) and (b).
There will be both resonant and off-resonant excitation if we use the rectangular pulse to drive electron spin \cite{scully1997quantum}.
The manipulation process can be described by Rabi oscillation \cite{rabi1937space} with Maxwell-Bloch equation
$\omega(t)= \omega(0)\{1+(\frac{\Omega'}{\Omega})^2[\cos (\Omega t) -1]\}$, where $\Omega'$ is determined by the driven microwave power, and $\Omega = \sqrt{\Omega^2+\Delta^2}$, $\Delta$ is the frequency detuning. When several transitions are needed to be manipulated simultaneously, there must be detuning for some of them. The oscillation amplitude will be small and the frequency asynchronous.
The flip becomes imperfect and the manipulation fidelity decreases drastically.
Though a increasing of the power is useful in some conditions, it may disturb other states need to be protected.

To overcome the aforementioned problems, BURP pulses are introduced.
They are a family of shaped pulses created by simulated annealing method
and standard gradient-descent routine to get near perfect quantum operations over prescribed frequency range,
while suppressing the rotation outside this region.
Because of an initial state independence property, application of the universal rotation by BURP pulses is of high efficiency in quantum computation with unknown initial states. Revealing most properties of shaped pulses, response of spins with various frequencies can be calculated by sequentially concatenating the quantum transformation within each time splices in the rotating frame.
As one typical BURP pulse,
REBURP 180 shape
(displayed in Fig.\ref{Shape}(c)),
has excellent self-refocusing behavior, which is revealed by the flat bottom of the response curve, and the sharply raised transition region as shown in Fig.\ref{Shape}(d).
The spins at the selective region can be inverted by the
shape with high fidelity, while the spins apart from the flat bottom are kept nearly untouched after the operation.
The narrow transition region, which bridges the flat bottom and the suppression region,
makes the REBURP shapes more applicable in practice, such as spin-echo experiments,
multidimensional spectroscopy and magnetic resonance imaging.

\section{Spin system of NV center}

%Characterize the system.
\begin{figure}
\includegraphics[width=\linewidth]{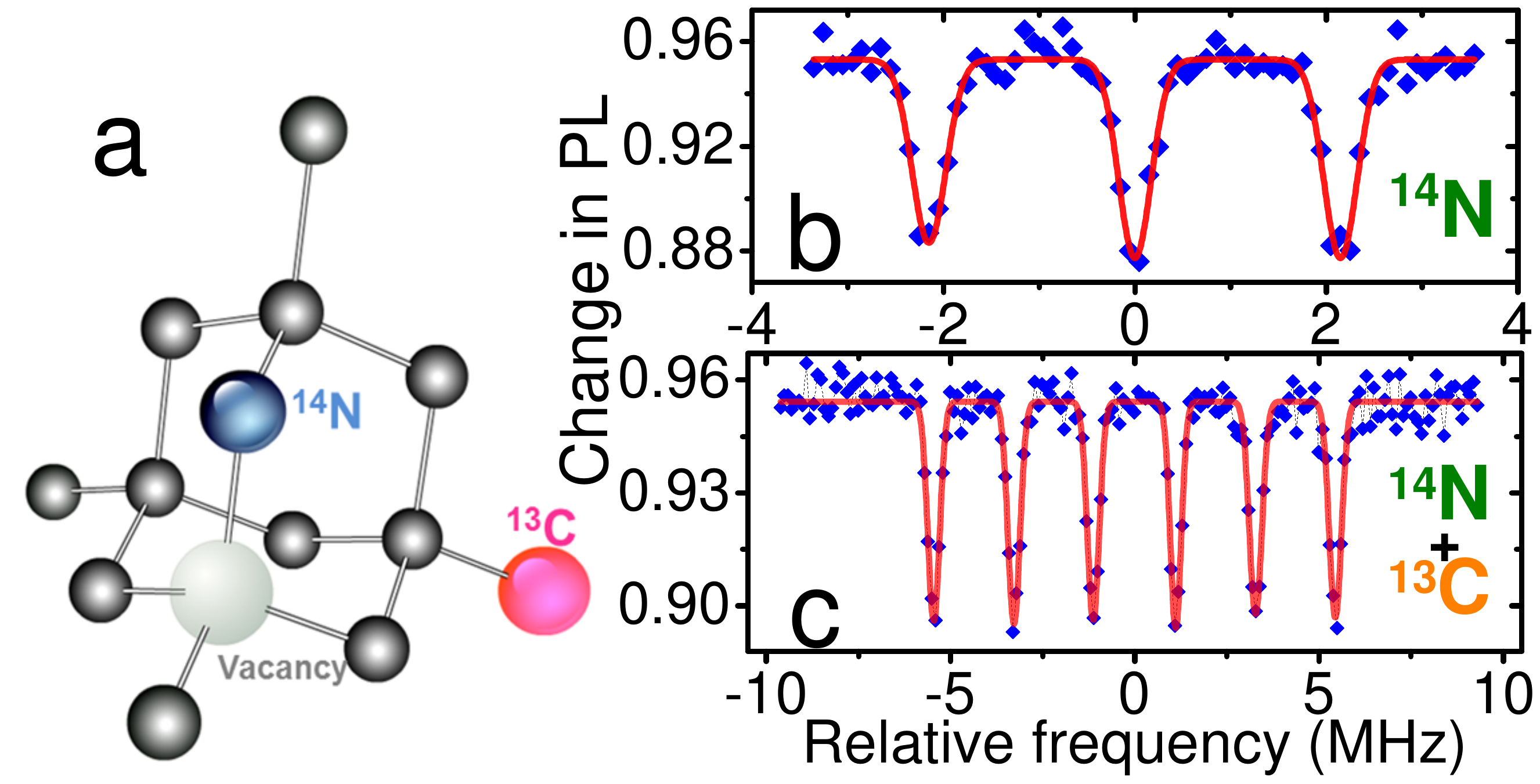} \caption{\label{Spectra}
\textbf{The structure and different ODMR spectra of NV center} (Color online).
(a) The structure of a single NV center in diamond.
(b) ODMR spectra of the NV center only with the host $^{14}N$ nucleus.
(c) ODMR spectra of the NV center with a $^{13}$C nucleus nearby.}
\end{figure}

To demonstrate the performance of the shaped pulses, we
implement experiment with one single NV center (structure shown in Fig.\ref{Spectra}(a)) in pure
diamond (Nitrogen concentration $<5~ppb$).
Lifting the electron spin degeneracy of $m_s=\pm1$
by external magnetic field $\mathbf{B}$,
the Hamiltonian of a negative charged NV center ($NV^-$) in pure diamond is:
%\begin{equation}\label{eq:H}
%\begin{split}
$H=\Delta S_{z}^2-\gamma_e \mathbf{B \cdot S}-\gamma_N\mathbf{B}\cdot \mathbf{I}^{(N)}-\gamma_C\mathbf{B}\cdot\sum_{i} \mathbf{I}_{i}^{(C)}+A_{\parallel}^{(N)} S_{z} I_{z}^{(N)}+A_{\perp}^{(N)} S_{x} I_{x}^{(N)}+A_{\perp}^{(N)} S_{y} I_{y}^{(N)}
+S_z\sum_{i}\mathbf{A}_{i}\cdot\mathbf{I}_{i}^{(C)}$
%\end{split}
%\end{equation}
where $\Delta=2.87~$ GHz is the zero-field splitting of the ground states.
%$\gamma_e=1.76\times10^{11} ~T^{-1}s^{-1}$  and $\gamma_c=6.73\times10^7 ~T^{-1}s^{-1}$
$\gamma_e$, $\gamma_C$ and $\gamma_N$ are the gyromagnetic ratio of electron spin and $^{13}C$, $^{14}N$ nuclear spins.
$\mathbf{A}_{i}$, $\mathbf{A}_{\parallel}^{(N)}$ and $\mathbf{A}_{\perp}^{(N)}$ are the hyperfine tensor for $^{13}C$ and $^{14}N$ nuclear spins.
Hyperfine spectra of the NV center is obtained by ODMR (optically detected magnetic resonance)\cite{van1988optically} scanning.
Fig.\ref{Spectra}(b) shows the hyperfine spectra induced by the interaction with the host $^{14}$N nucleus. The electron spin
couples with the $I_{N}=1$
nuclear spin, thus m$_{s}=-1$ level splits into 3 energy levels.
Each dip in the spectra represents one allowed transition related to the host $^{14}N$ nuclear spin state.
When there are also an ambient $^{13}$C nucleus %coupling with the electron spin,
, the hyperfine spectra
are displayed in Fig.\ref{Spectra}(c).
The interaction with a
$^{13}$C atom nearby (
$I_{C}=\frac{1}{2}$ )
splits the 3 energy levels mentioned above into 6 levels, causing 6 peaks in the spectra.
More other ambient $^{13}$C nuclei can be also coupling to the NV electron spin,
and the abundant hyperfine structure can serves as multi-qubit resources.

\section{Experiment}

%\subsection{Experiment setups}

A home-built confocal microscope system with microwave (MW) components is used to
initialize, manipulate and readout the spin state of the NV center.
To enhance the photon collection efficiency, a solid immersion lens (SIL)\cite{marseglia2011nanofabricated} is etched above an NV center.
Because the NV center investigated is one pure center without ambient $^{13}C$,
different hyperfine spectra occasion are simulated by adjusting the external permanent magnet.
Owning to the pure phase property of the BURP shapes,
only one microwave source and path are required to supply the carrier frequency MW and to transfer the MW signal.
Both the rectangular and shaped MW pulses are acquired by modulating the carrier frequency MW using an arbitrary waveform generator (Tektronix AWG430) through mixers. After being amplified, the MW pulses are delivered to the NV center through a coplanar waveguide (CPW) antenna deposited close to the SIL.
All these experiments are done at room temperature.

\subsection{Frequency-sweep experiment}

\begin{figure}
\includegraphics[width=\linewidth]{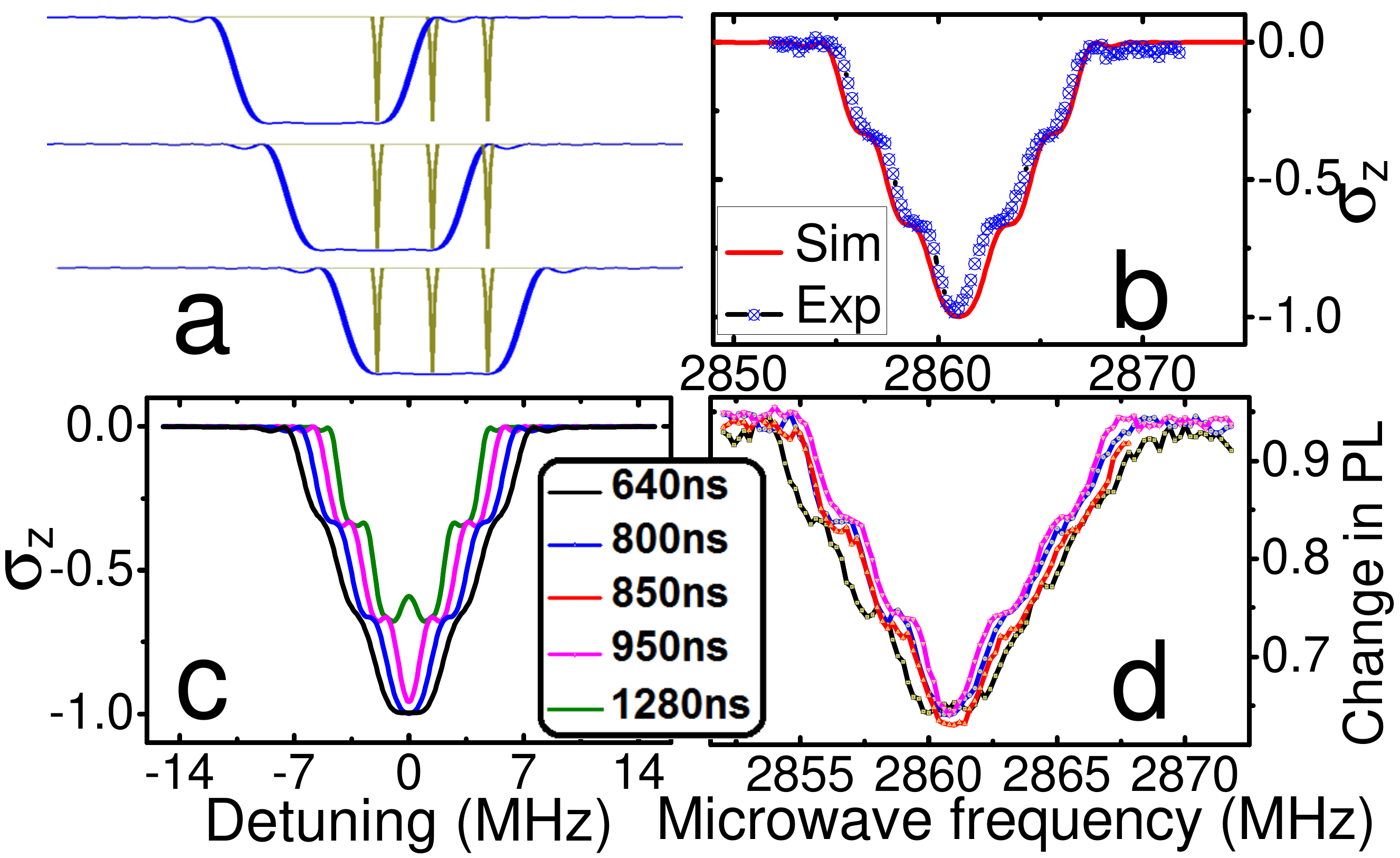} \caption{\label{Sweep}
\textbf{Spin response results of the frequency-sweep process} (Color online).
(a) Frequency sweep  procedure across hyperfine spectra (with 3 dips of the host $^{14}N$) driven by the REBURP 180 pulses.
(b) Comparison of the simulation and experimental results about one given timescale ($800 ns$) pulse.
(c) Simulation results of different timescales ($640 ns, 800 ns, 950 ns, 1280 ns$) pulses.
(d)Experimental results of different timescales ($640 ns, 800 ns, 850 ns, 950 ns$) pulses.}
\end{figure}

In order to verify the excellent self-refocusing behavior of the REBURP 180 pulse,
we investigate the spin response results by frequency-sweep experiment
across the $^{14}N$ hyperfine spectra (Fig.\ref{Spectra}(b)).
The experiment is similar to pulse ODMR \cite{jelezko2004read} scanning,
but only replace rectangular $\pi$ pulse with the REBURP 180 pulse.
One adjustable parameter of the pulse is timescale
corresponding to different peak power used in the pulse.
We choose one $800 ns$ timescale pulse,
of which the flat bottom in spin response curve (Fig.\ref{Shape}(d)) can just cover all 3 dips in the spectra.
Simulating the the frequency-sweep procedure,
we can see that the 3 transitions are driven one by one, Fig.\ref{Sweep}(a).
Rather than Gauss-like profile in traditional Pulse ODMR,
a stepped decreasing and followed increasing response profile emerges in Fig.\ref{Sweep}(b).
In comparison with simulation, the experiment results are shown in Fig.\ref{Sweep}(b).
Steps in the profile indicate that
one or two of the transitions can be completely excited with others untouched,
corresponding to the sharped edges in spin response curve of the REBURP pulse (Fig.\ref{Shape}(d)).
The central dip with triple depth of one step indicates that
the three transitions can be excited perfectly simultaneously,
corresponding to the wide and flat bottom in spin response curve.
We can get different band-width pulses by changing the pulse timescale,
with their simulation and experimental results for frequency-sweep experiment shown in Fig.\ref{Sweep}(c) and (d).
The self-refocusing behavior of the REBURP pulse is confirmed by the consistence between experiment and simulation.

\subsection{Multi-flip experiment}

\begin{figure}
\includegraphics[width=\linewidth]{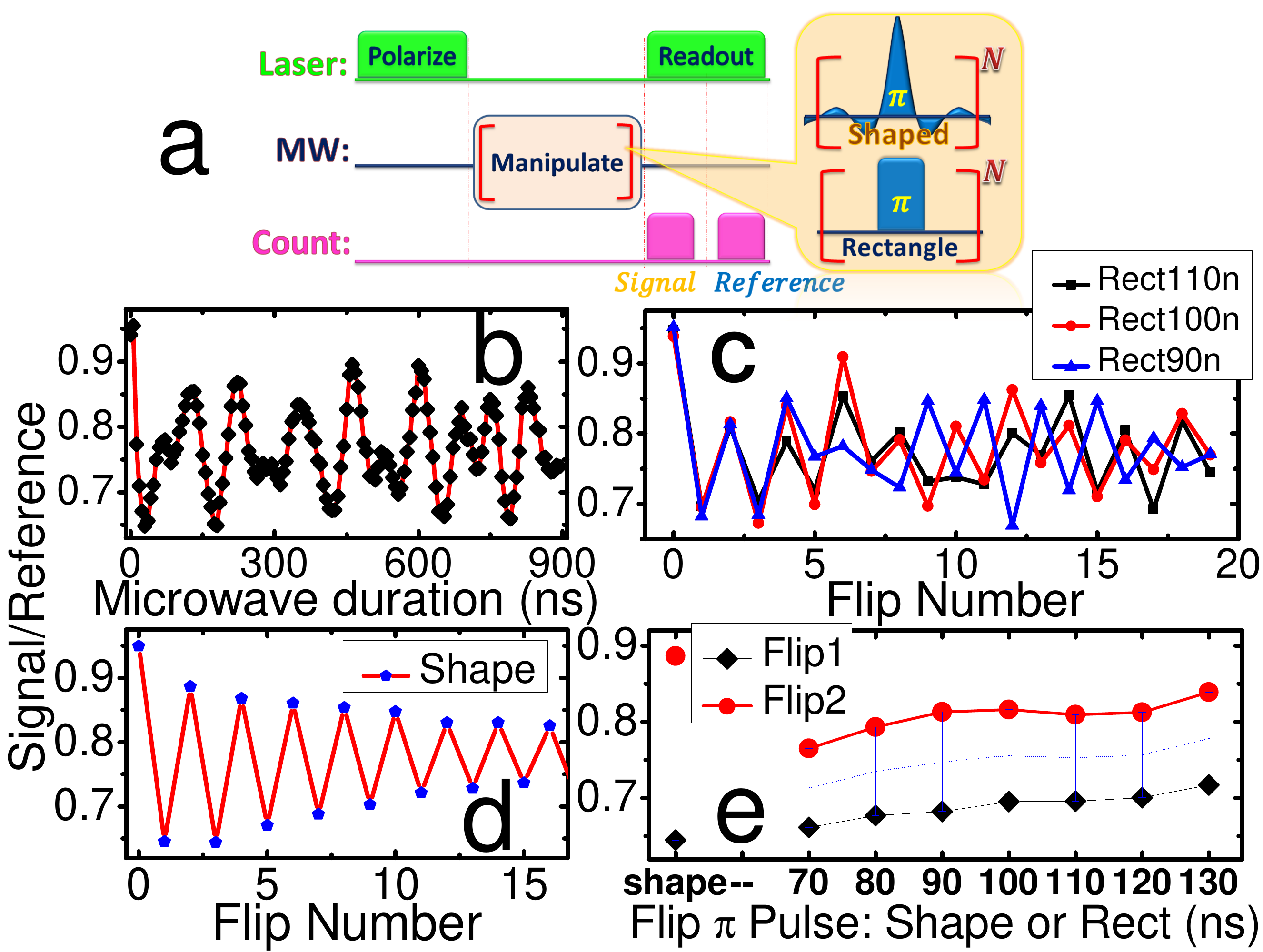} \caption{\label{Flip}
\textbf{Multi-flip effects of rectangular and shaped pulse}(Color online).
(a)The sequence scheme of multi-flip experiment.
(b) Serious beating in Rabi oscillation driven by rectangle pulses.
(c) Multi flip manipulated by different timescale ($110 ns, 100 ns, 90 ns$) rectangle $\pi$ pulses.
(d) Multi flip operated by RENURP 180 shaped pulses.
(e) Comparison between the shaped pulse and different timescale rectangle pulses
    about the amplitude during the second flip (difference of Flip1 and Flip2). }
\end{figure}

To further test the control fidelity improvement of the shaped pulse,
multi-flip experiment is employed under ODMR spectra with 6 contiguous dips (Fig.\ref{Spectra}(c)).
This typical circumstance is to simulate $6.5 MHz$ level spilling induced by a $^{13}C$ nucleus.
The experiment target is to flip the three levels in the left of the spectra repeatedly by rectangular or shaped pulses, with the right ones remain intact. The spin states after increasing number of flips are all readout with sequence scheme shown in Fig.\ref{Flip}(a).
Several Rabi oscillation driven by rectangle pulses
are implemented to determine rectangular $\pi$ pulses,
one of them shown in Fig.\ref{Flip}(b).
However, there are serious beatings caused by
off-excitations with different detunings for different energy level transitions.
The wide sinc-like spectrum of the rectangular pulses in spin response curve (Fig.\ref{Shape}(b))
make it impossible to drive the left three energy levels synchronously without disturbing the right ones.
Multi-flip test using several timescale rectangular $\pi$ pulse are carried out,
and the curves become totally a mess within only a few flips,
with three better ones shown in Fig.\ref{Flip}(c).
In contrast, multi-flip by REBURP 180 shaped pulse is also implemented,
and it can last at least 16 flips, indicating a far better fidelity than rectangular pulses in Fig.\ref{Flip}(d).
By the difference of states after once flip (Flip1) and twice flips (Flip2),
we can evaluate the efficiency of the second flip, and furthermore estimate effects of multi flips.
Of the flip amplitudes during Flip1 and Flip2,
a comparison is made between shaped pulse and different timescale rectangular $\pi$ pulses in Fig.\ref{Flip}(e).
Obviously, the flip amplitude of whichever timescale rectangular $\pi$ pulse
are much lower than (about half of) that of the shaped pulse.
From advantages in flip numbers and amplitudes,
we can see that the control fidelity is improved more than twice by the shaped pulse,
compared with the traditional rectangular pulses.

\section{Conclusions}

In summary, we have introduced a technique, REBURP shaped pulse
to improve the control fidelity,
which is verified by experiments as well as simulation, in NV center spin system.
The REBURP pulse has many excellent properties,
such as self-focusing, band-selective, uniform response and pure-phase.
It can have almost equal excitation effect among a wide-band but sharply edged frequency region,
while redundant off-resonant excitations out the region
are well suppressed.
The spin response characters of the REBURP pulse are confirmed by frequency-sweep experiments.
The repeatedly flip numbers and multi-flip amplitudes of the shaped pulse
both have great advantages over traditional rectangular pulses in the mentioned typical circumstance.
The control fidelity can be improved more than twice by the shaped pulse, compared with rectangular pulses.
Because of the pure phase property, only one microwave resource and path are required,
which make the application of BURP pulses more robust and easy access.
Not only in experiments of NV centers in diamond,
the approach introduced in this paper can also applied to NMR or other quantum systems
that require high fidelity band-selective operations.

\begin{acknowledgments}

 This work is supported by the National Basic Research Program of China (973 Program under Grants Nos. 2014CB921402 and 2011CB9216002), the National Natural Science Foundation of China (under Grants Nos. 11175094, 91221205, 91123004, 51272278, and 91023041). GLL also thanks the support of Center of Atomic and Molecular Nanoscience of Tsinghua University.

\end{acknowledgments}

% \emph{Authors' Note}: During the preparation of this manuscript, we got aware of a similar
%investigation \cite{Jelezko DNP} in which the Hartmann-Hahn
%condition was used to investigate the  $^{13}$C DNP in large
%magnetic fields. We thank Fedor Jelezko for the helpful discussion.

%\input acknowledgement.tex   % input acknowledgement

\end{document}